# Accelerating R with high performance linear algebra libraries

**Bogdan Oancea**

"Nicolae Titulescu" University of Bucharest and National Statistics Institute of Romania,

bogdanoancea@univnt.ro

**Tudorel Andrei**

The Bucharest University of Economic Studies,

andreitudorel@yahoo.com

**Raluca Mariana Dragoescu**

The Bucharest University of Economic Studies

dragoescuraluca@gmail.com

**Abstract**

Linear algebra routines are basic building blocks for the statistical software. In this paper we analyzed how can we can improve R performance for matrix computations. We benchmarked few matrix operations using the standard linear algebra libraries included in the R distribution and high performance libraries like OpenBLAS, GotoBLAS and MKL. Our tests showed the the best results are obtained with the MKL library, the other two libraries having similar performances, but lower than MKL.

**Keywords**: R, linear algebra, BLAS, high performance computing.

**1. Introduction**

Linear algebra routines are essential building blocks for statistical software. Statistical and data analysis software use operations like matrix multiplication, matrix inversion, LU or Cholesky factorizations, QR or SVD decompositions, eigenvalue or eigenvector decompositions.

Many times, the data sets or the matrices being processed are very large. When processing very large data sets a problem could rise: can R be used to process such large amounts of data? R loads and processes every data set in memory and this can be a serious limitation on the size of the data set. More, R is an interpreted language (like Matlab or Octave) and interpreted languages are known to be slower than direct execution of the code generated by programs written in languages like C or C++. Given these disadvantages one can ask if R is appropriate software for processing very large matrices. The answer to this question is: yes, R could be used for very large matrices and we will try to show how can R could obtain high performance in matrix computations.

R has some advantages over other high performance languages: easy access to the data, easy visualization of the data, a lot packages for statistical computing etc. Besides, R uses linear algebra libraries for matrix computations: BLAS (Blackford et. al. 2002) and LAPACK (Anderson et al., 1999). The fact that R uses specialized libraries for matrix computations can be exploited to achieve high performance.

Accelerating R for high performance linear algebra computations can be achieved by means of:

- Developing specialized libraries for matrix computations in C++ and interfacing them with R;
- Using high performance linear algebra libraries instead of the standard libraries provided by the R distribution;
- Using multicore or GPU libraries;

## 2. Literature review

At this moment there are only few studies regarding high performance computations in R using the approaches that we've mention in introduction.

The first approach, interfacing specialized libraries developed in C++ with R is treated in a number of papers. Bates and Eddelbuettel (2013) describe a package called RcppEigen that provides access from R to Eigen C++ template library for numerical linear algebra. The package described by the authors provides an interface to pass object from R to C++ and back. This allows R users to obtain near C++ performance for matrix computations. In another paper, Eddelbuettel and Sanderson (2014) presents RcppArmadillo package that provides R and interface to the C++ Armadillo matrix library. Armadillo is a C++ linear algebra library, with a good balance between speed and ease of use. It provides classes for vectors, matrices and cubes, as well as functions that operate on these classes. By means of RcppArmadillo R users can access the functionality of the Armadillo library.

The second approach to improve the performance of R is to using high performance linear algebra libraries instead of the standard libraries provided by the R distribution. This approach has been addressed in a number of studies. Rodrigues (2014) presents a benchmark of few matrix operations using plain R, OpenBLAS (OpenBLAS, 2015), ATLAS (Whaley, 1999) and MKL (Intel, 2014) libraries. He shows that R with specialized libraries can run about 4 times faster that R with the standard libraries.

In another work, de Vries (2014) presents the improvements obtained by using MKL library in Revolution R Open, showing that a matrix multiplication can run 27 times faster than standard R, a linear regression 20 times faster, a Cholesky factorization 17 times faster, a matrix inversion 10 times faster and a discriminant analysis 3.6 times faster.

Riedel (2012) presents a benchmark of linear algebra operations using the default libraries from R, GotoBLAS (Goto, 2008) and MKL libraries, showing a speedup of 5-7 times for matrix multiplication and 10-12 times for SVD.

**3. Experimental setup**

In order to test the performance of matrix computations in R we run a series of experiments using different high performance linear algebra libraries and the standard libraries from the R distribution. In our experiments we used the following linear algebra libraries:

- Standard BLAS and LAPACK libraries, included in R distribution;

- GotoBLAS library (Goto, 2008) – it was developed by Kazushige Goto at the Texas Advanced Computing Center during 2002 and it is freely available since 2005. It contains handcrafted assembly code to obtain high performance. Unfortunately it is no longer developed since 2008;

- OpenBLAS library (OpenBLAS, 2015) - is a continuation of the GotoBLAS. It adds optimized implementations of linear algebra kernels for several processor architectures, including Intel Sandy Bridge and Loongson;

- MKL library (Intel, 2014) – is a library developed at Intel that contains routines for scientific computing including BLAS and LAPACK, sparse solvers and Fast Fourier transforms functionalities, highly optimized for Intel processors. We tested this library by using Revolution Open R that includes MKL.

All tests were run using R 3.1.3 and Revolution R Open 8.0.1 under Windows 7 - 64 bit operating system. The computer used for benchmark has a Intel i5-2410M processor at 2.3 GHZ with 2 cores, 128 L1, 512kB L2 and 3 MB L3 cache and 4 GB DDR3 RAM.

There are two ways of using a different linear algebra library instead of the one provided with the standard R distribution. One way is to compile the linear algebra library as a static library and link it with R. This would mean to recompile R. The advantage of this approach is that it will run faster because a static library is loaded into main memory together with the main program and it will not incur any overhead during execution. There is also a disadvantage: it is not an easy task to recompile R and link it against the static linear algebra libraries, at least for an R user that is not a computer programmer. The second option involves compiling the linear algebra as a dynamic library (.dll under Windows or .so under Linux) and replaces the library in the R installation (Rblas.dll and Rlapack.dll under

Windows, libRblas.so under Linux). This option can be done even by an inexperienced R user. The disadvantage is that it will run slower because a dynamic library is loaded in memory only on demand, when the execution thread calls a function from the library. This will incur some overhead, but the overhead can be reduced by repeating the execution of the same function many times and take an average running time.

Considering the advantages and disadvantages of both approaches mentioned above we chose the second approach and we benchmarked the following routines:

- Matrix multiplication;
- Matrix inversion;
- QR decomposition;
- LU factorization;
- Cholesky factorization;
- SVD decomposition.

For our tests we used matrices with 1000, 2000, 3000, 4000, 5000 and 6000 rows/columns.

## 4. The results

For each test we generated a random matrix and used the Matrix package (Bates, 2010) for representing a matrix object. We run each operation 3 times, recorded the running time using system.time() function and computed an average time. We than computed the speedup as the number of times an operations run faster using a high performance linear algebra library than using the standard library included in the R distribution. Below are the results of the tests.

Table 1. OpenBLAS speedup

| Matrix dimension | Matrix Multiplication | Matrix Inversion | QR | LU | Cholesky | SVD |
|---|---|---|---|---|---|---|
| 1000 | 7.4 | 3.4 | 1.7 | 2.0 | 3.3 | 2.3 |
| 2000 | 10.6 | 5.1 | 1.8 | 4.0 | 5.6 | 2.7 |
| 3000 | 10.9 | 5.5 | 1.8 | 4.8 | 6.7 | 2.9 |
| 4000 | 11.4 | 5.4 | 1.9 | 5.1 | 6.8 | 2.9 |
| 5000 | 11.5 | 5.6 | 1.6 | 6.3 | 6.9 | 2.9 |
| 6000 | 9.8 | 4.9 | 1.8 | 5.5 | 6.6 | 3.0 |

Table 2. GotoBLAS speedup

| Matrix dimension | Matrix Multiplication | Matrix Inversion | QR | LU | Cholesky | SVD |
|---|---|---|---|---|---|---|
| 1000 | 7.7 | 3.2 | 1.8 | 2.7 | 2.2 | 2.4 |
| 2000 | 8.9 | 5.0 | 1.9 | 3.5 | 6.0 | 3.2 |
| 3000 | 9.2 | 5.8 | 2.0 | 4.2 | 5.9 | 3.3 |
| 4000 | 9.7 | 6.1 | 2.0 | 4.0 | 6.1 | 3.3 |
| 5000 | 9.1 | 6.0 | 2.0 | 5.0 | 6.3 | 3.4 |
| 6000 | 9.4 | 6.4 | 2.0 | 4.7 | 6.6 | 3.2 |

Table 3. MKL speedup

| Matrix dimension | Matrix Multiplication | Matrix Inversion | QR | LU | Cholesky | SVD |
|---|---|---|---|---|---|---|
| 1000 | 14.2 | 2.3 | 2.0 | 7.3 | 6.7 | 3.3 |
| 2000 | 15.9 | 10.7 | 2.0 | 7.6 | 13.7 | 4.0 |
| 3000 | 17.3 | 11.1 | 2.1 | 7.1 | 14.1 | 4.1 |
| 4000 | 17.2 | 11.5 | 2.1 | 8.5 | 11.7 | 4.2 |
| 5000 | 17.3 | 11.9 | 2.1 | 9.7 | 12.8 | 4.5 |
| 6000 | 14.5 | 10.4 | 2.1 | 5.6 | 5.2 | 3.8 |

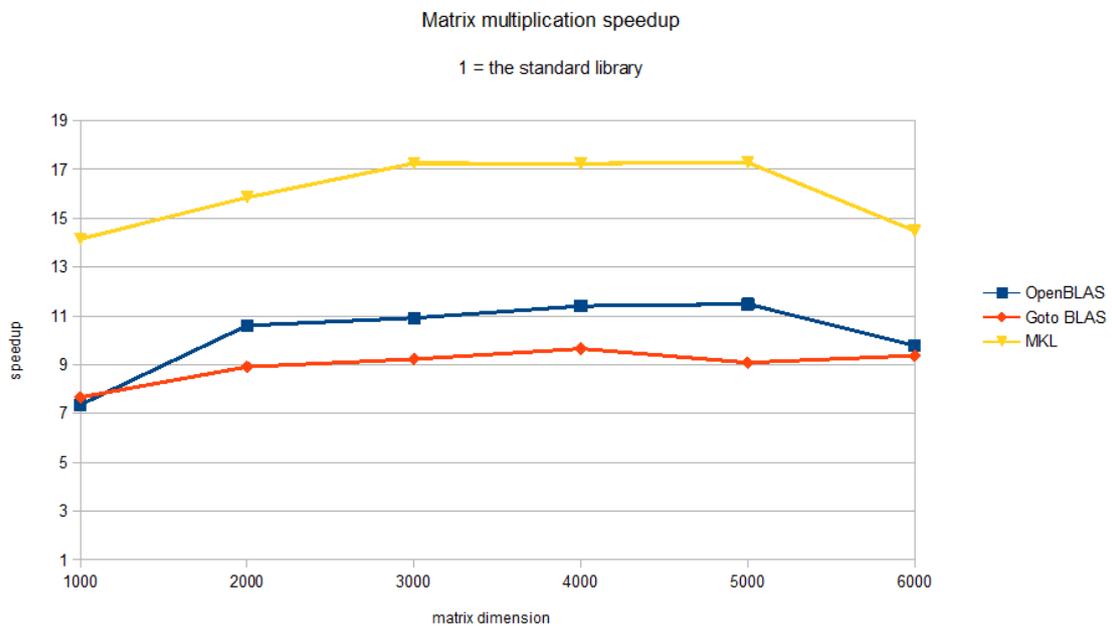

Figure 1. Matrix multiplication speedup

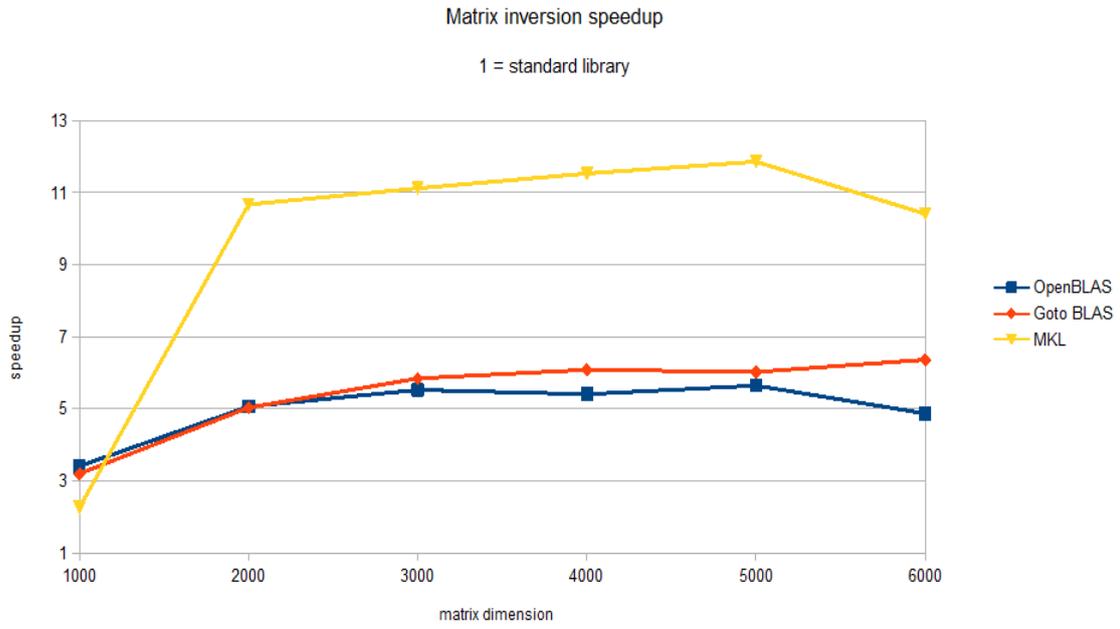

Figure 2. Matrix inversion speedup

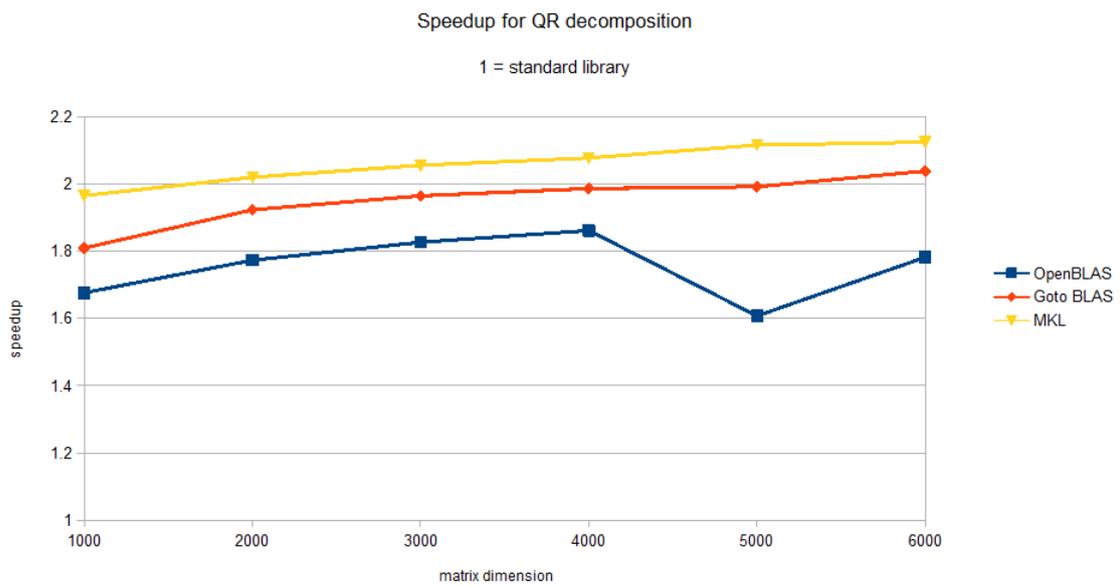

Figure 3. QR speedup

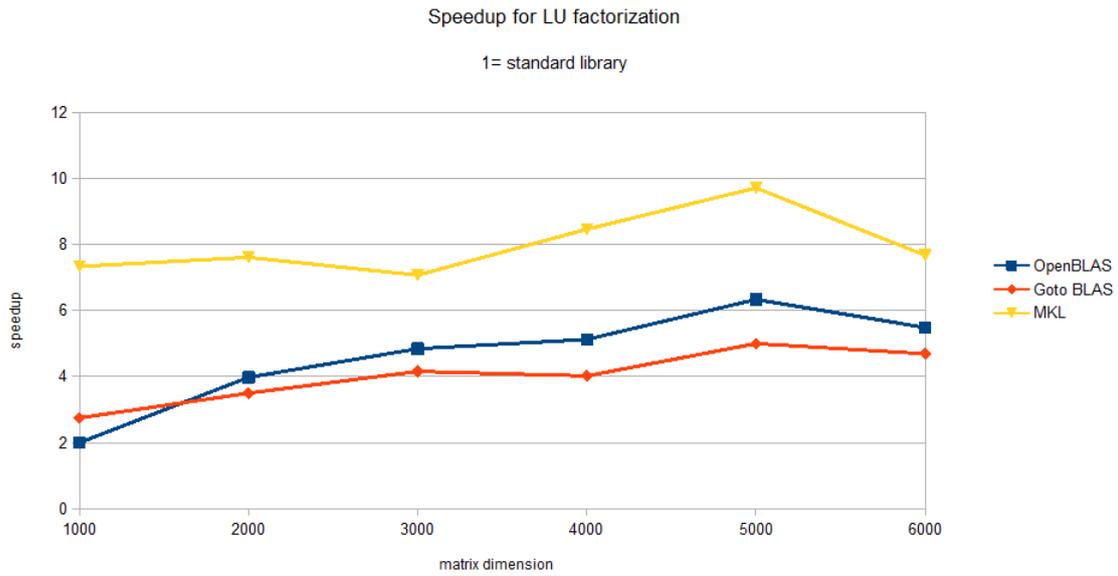

Figure 4. LU factorization speedup

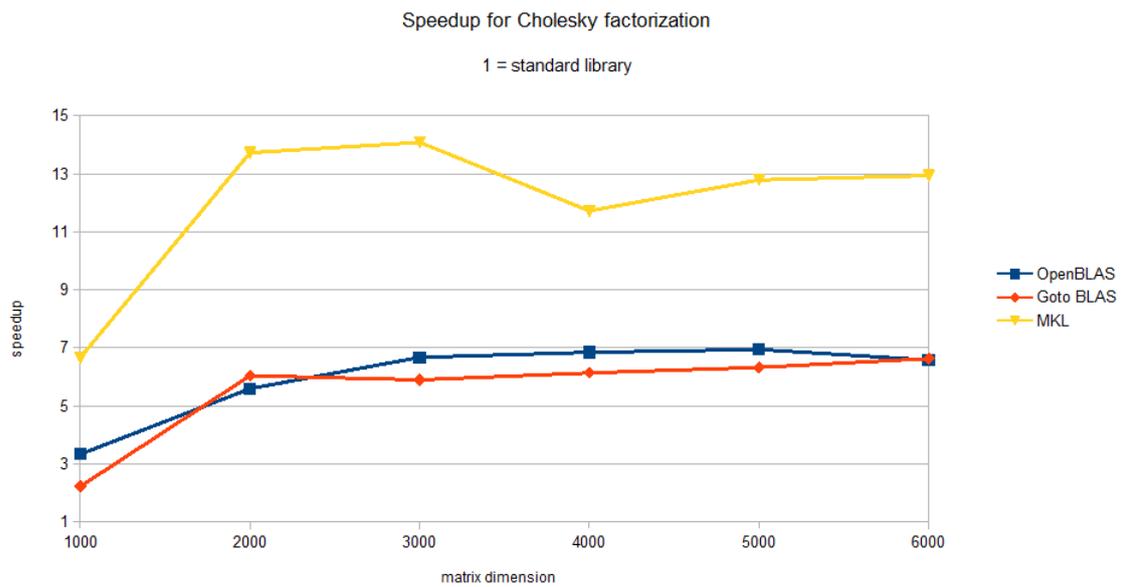

Figure 5. Chlesky factorization speedup

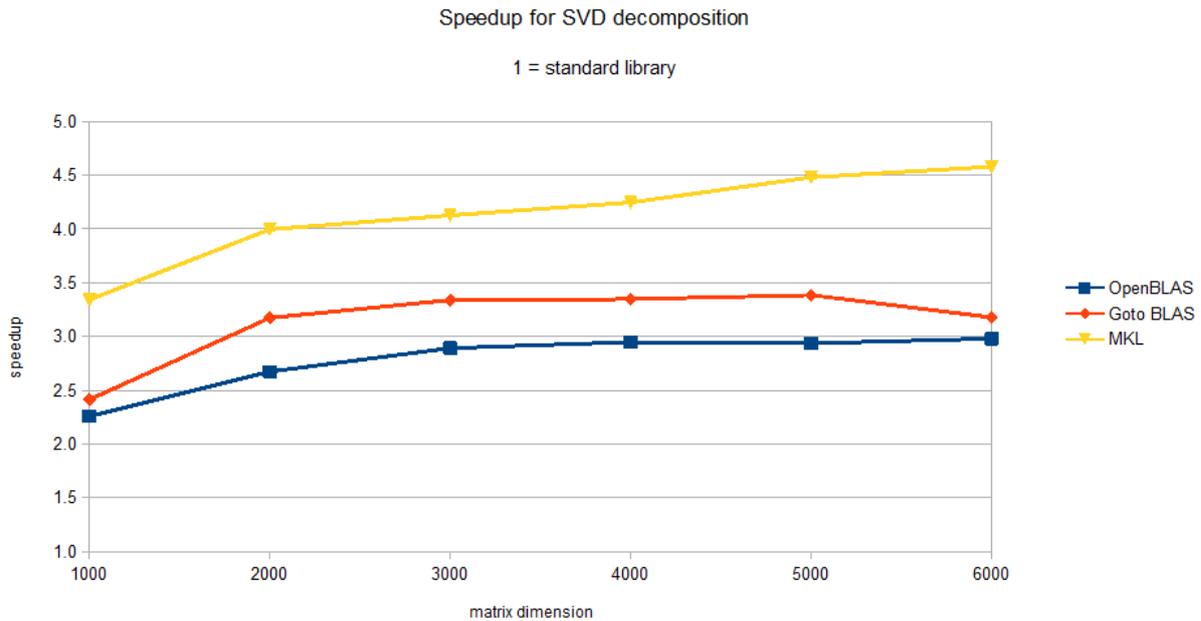

Figure 6. SVD speedup

## 5. Conclusions and future work

Analyzing the results from table 1 to 3 and figures 1 to 6 we can draw the conclusion that MKL library outperforms GotoBLAS and OpenBLAS. The largest speedup was obtained for matrix multiplication (MKL ~ 17, GotoBLAS ~9 OpenBLAS ~10). GotoBLAS and OpenBLAS have relatively similar performance (GotoBLAS has a better performance for matrix inversion, QR and SVD decomposition while OpenBLAS performs better for matrix multiplication, LU and Cholesky factorizations) but lower than MKL.

These tests showed that replacing the standard BLAS and LAPACK libraries could significantly improve R performance for linear algebra computations and R users can obtain performances similar to other direct compiled programs.

As a future work we intend to extend our test using ATLAS library and multithreaded versions of the linear algebra libraries and also test libraries that combine GPU and CPU for computations. We also intend to compare R performance with other scientific software packages (Octave, Matlab etc).

**Acknowledgment**

This paper was co-financed from the European Social Fund, through the Sectorial Operational Programme Human Resources Development 2007-2013, project number POSDRU/159/1.5/S/138907 "Excellence in scientific interdisciplinary research, doctoral and postdoctoral, in the economic, social and medical fields -EXCELIS", coordinator The Bucharest University of Economic Studies.